\documentclass[12pt,preprint]{aastex}

\newcommand{\htwo}{H\,{\scriptsize\sc{II}} }

\begin{document}

 \title{Search for Calibrators for the Submillimeter Array:
       I. High-Mass Star Forming Regions}

 \author{Y.-N. Su\altaffilmark{1,2}, S.-Y. Liu\altaffilmark{1}, J.~Lim\altaffilmark{1}, N.~Ohashi\altaffilmark{3}, H.~Beuther\altaffilmark{4}, Q.~Zhang\altaffilmark{4}, P.~Sollins\altaffilmark{4}, T.~Hunter\altaffilmark{4}, T. K. Sridharan\altaffilmark{4}, J.-H.~Zhao\altaffilmark{4}, and
 P.~T.~P.~Ho\altaffilmark{1,4}}

 \altaffiltext{1}{Institute of Astronomy and Astrophysics, Academia Sinica, P.O. Box 23-141, Taipei 106, Taiwan, R.O.C.}
 \altaffiltext{2}{Institute of Astronomy, National Central University, 300 Jungda Rd, Jungli City, Taoyuan 320, Taiwan, R.O.C.}
 \altaffiltext{3}{Institute of Astronomy and Astrophysics, Academia Sinica, 645 North A'ohoku Place, Hilo, HI 96720}
 \altaffiltext{4}{Harvard-Smithsonian Center for Astrophysics, 60 Garden Street, Cambridge, MA 02138}

\begin{abstract}
We present initial results of an ongoing search for
interferometric calibrators at submillimeter (sub-mm) wavelengths
with the Submillimeter Array (SMA). Powerful radio galaxies are
commonly used as calibrators at centimeter and millimeter
wavelengths, but many are not strong enough to serve as
calibrators at sub-mm wavelengths because of their rapidly
declining flux densities toward shorter wavelengths. The inability
to find a calibrator close to the target source may limit or even
prevent us from imaging many interesting sources at sub-mm
wavelengths. Here, we investigate whether high-mass protostellar
objects and ultracompact \htwo regions can serve as useful
calibrators for the SMA. The dust emission associated with these
objects makes them among the brightest sub-mm sources in the sky.
Our observations at 0.85~mm (345 GHz) with an angular resolution
of $\sim$3$\arcsec$ reveal that although a large fraction of the
dust emission originates from an extended ``halo'' component, a
compact unresolved component often remains that when sufficiently
strong may serve as a useful calibrator. These observations also
provide a first glimpse at the small-scale distribution of dust
around ultracompact \htwo regions and high-mass protostellar
objects at sub-mm wavelengths. We discuss the origin of the
core-halo structure seen in many sources, and conclude with
suggestions for future searches for calibrators with the SMA.
\end{abstract}


\section{Introduction}
Calibrators used for determining the amplitude and phase
(hereafter referred to as gain) of an interferometer are usually
compact (i.e., spatially unresolved) and bright. The difference
between the measured and expected gain of the calibrator,
introduced by the Earth's atmosphere and telescope electronics,
are used to correct the gain measured towards the target source.
Because transmission (affecting the amplitude) and refraction
(affecting the phase) through the Earth's atmosphere can differ
significantly along different lines of sight, the calibrator
should lie close in the sky to the target source. At centimeter
and millimeter wavelengths, powerful radio galaxies are
sufficiently compact, bright and numerous to serve as calibrators
in virtually any region of the sky. At submillimeter (sub-mm)
wavelengths, however, most of these radio galaxies are too weak to
serve as calibrators for the Submillimeter Array\footnote{%
The Submillimeter Array is a joint project between the Smithsonian
Astrophysical Observatory and the Academia Sinica Institute of
Astronomy and Astrophysics, and is funded by the Smithsonian
Institution and the Academia Sinica.} (SMA). This is because their
synchrotron emission usually decreases rapidly in intensity
towards shorter wavelengths, and furthermore the instrumental
noise is higher because of higher system temperatures at shorter
wavelengths. We estimate that only about eighty radio galaxies are
bright enough to serve as calibrators for the SMA at 0.85 mm (345
GHz), so that the mean angular separation between calibrators in
the sky is $\sim$20$^\circ$. This may limit our ability to image
many interesting sources, especially at high angular resolutions
where accurate gain corrections (and hence close proximity of a
calibrator) become increasingly important. At 0.45 mm (690 GHz)
the situation is drastically worse, with only about a dozen radio
galaxies useable as calibrators for the SMA; none are located at
R.A. 5$^h$$-$19$^h$ and Dec. $>$ 5$^\circ$. Even with the addition
of planets and their moons to the list of available calibrators,
we still may not be able to image sources in large swaths of the
sky.

By contrast with synchrotron emission, thermal emission from dust
has a rapidly increasing intensity towards shorter wavelengths, so
that dust sources are good candidates for calibrators at sub-mm
wavelengths. A particularly promising class of objects are young
high-mass stars, which can often be found embedded in their natal
molecular cloud cores even when they have produced sufficient UV
photons to excite an ultracompact (UC) \htwo region. The emission
from their dust cocoons makes them among the brightest sources in
the Galaxy at sub-mm wavelengths; in single-dish observations,
high-mass star-forming regions can exhibit continuum emission of
up to a few hundred Jy at 0.35 mm \citep{san94,hun00}. Their dust
distribution on the relative small spatial scales that
interferometers probe, however, is not known. Recently, high
angular resolution ($\sim$0\farcs5) mid-IR observations of the UC
\htwo region G29.96$-$0.02 showed a good spatial coincidence
between warm dust grains and ionized gas imaged at 1.3 cm
\citep{deb02}. If the cooler dust probed at sub-mm wavelengths has
a similar distribution, UC \htwo regions may make suitable
calibrators at sub-mm wavelengths. Their location in the Galactic
plane makes them an especially important addition to the available
calibrators for studies of star-forming regions and other Galactic
objects. In this Letter, we present the first results at 0.85 mm
to determine whether high-mass star forming regions can serve as
calibrators for the SMA. Because of the relative paucity of
available calibrators at 0.45 mm, this also is a necessary first
step towards determining whether young high-mass stars can
potentially serve as calibrators at shorter wavelengths.

\section{Observations}
\subsection{Sample Selection and Observational Strategy}
To preferentially select for point-like dust emission, we targeted
eight high-mass star forming regions that are either UC \htwo
regions with very small angular sizes ($\lesssim$1$\arcsec$) as
cataloged by \citet{woo89} and \citet{kur94}, or high-mass
protostellar objects (HMPOs) with a centrally peaked morphology in
mm (dust) continuum as compiled by \citet{sri02} and
\citet{beu02}. The five selected UC \htwo regions are relatively
isolated, with no other known compact ionized gas sources within
the $\sim$35$\arcsec$ field of view of the SMA thereby avoiding
confusion from nearby young massive stars. The three selected
HMPOs are each associated with a centrally peaked 1.2-mm (dust)
continuum source imaged with the IRAM 30-m telescope at an angular
resolution of 11$\arcsec$ \citep{beu02}. Follow-up interferometric
observations at higher angular resolutions (2\arcsec$-$4\arcsec)
revealed just a compact source in each case
\citep{beuther03a,beuther03b,wyr03}. The properties of the target
sources are summarized in Table~1.

\subsection{Observational Parameters}
Observations in the 0.85 mm band were carried out with the SMA
between 2003 May and 2003 September in its compact and extended
configurations with 4 antennas in each configuration. The combined
arrays provided projected baselines ranging from approximately
10~m to 120~m. To optimize the {\it uv} coverage, 2$-$3 sources
were observed alternatively in a track with an on-source
integration time of $\sim$1 hour per source. The half-power width
of the SMA primary beam at 0.85 mm is $\sim$35$\arcsec$, and the
total available double-sideband bandwidth was 2 GHz at the time of
the observations. See \citet{ho03} for more complete
specifications of the SMA. The absolute flux density scale was
determined from observations of either Uranus or Neptune, and is
estimated to have an uncertainty of $\sim$25\%. The nearest
available radio galaxy was observed as the gain calibrator for
each target source. We calibrated the data using the MIR software
package adapted for the SMA from the MMA software package
developed originally for the OVRO \citep{sco93}. We made maps
using the MIRIAD package. With natural weighting, typical angular
resolutions were about 3$\arcsec$$-$4$\arcsec$ and noise levels
were 30 mJy beam$^{-1}$.

\section{Results and Discussion}
\subsection{Structure of the 0.85 mm Continuum Emission}
We detected continuum emission towards all target sources. The two
sources G10.841$-$2.592 and Onsala~1 show distinct multiple
components, as shown in the left panel of Figure 1 for Onsala 1.
These sources are for the time being ruled out as suitable
calibrators at 0.85 mm. For the remaining six sources, five show
what appears to be a single compact component, whereas in the case
of G15.04$-$0.68 extended emission is also detected. As
diffuse/extended structures can be difficult to discern in the
image plane, we also examined the amplitude of the measured
visibilities as a function of projected baseline length (i.e.,
{\it uv} distance). For an extended source, the visibility
amplitude generally decreases with increasing {\it uv} distance,
whereas a point source shows a constant visibility amplitude over
all {\it uv} range. As an example, the right panel of Figure~1
shows a map of the 0.85 mm continuum towards G45.07+0.13. This
image reveals a compact ``core'' component that is at best only
marginally resolved. The visibility amplitude plot of
Figure~\ref{g45-uv}, however, clearly reveals an extended ``halo''
structure. Indeed, all the five sources that appeared to be
compact in our maps were found to exhibit extended emission.

To estimate the flux densities and sizes of the different
components in each target source, we first fitted the measured
visibilities with a one-dimensional Gaussian source structure
(i.e., halo component). When this fit proved unsatisfactory, we
added a point source (i.e., core component). As shown in Figure
\ref{g45-uv}, the visibility amplitudes of G45.07+0.13 can be fit
with a core-halo structure with flux densities of 0.90 Jy and 2.71
Jy for the two respective components, and a half-power width for
the halo of $\sim$3\farcs8. The results of the fits for all six
sources are listed in Table~2. Two are satisfactorily fit by a
single Gaussian component, whereas the four sources G15.04$-$0.68,
G45.07+0.13, IRAS 18089$-$1732, and IRAS 19217+1651 require a core
component with flux density of at least 0.35 Jy in addition to a
halo component.

\subsection{Suitability as Calibrators}
The first two sub-mm bands at which the SMA can operate are 0.85
and 0.45 mm. With a system temperature of 500/2000 K and aperture
efficiency of 0.70/0.60 at 0.85/0.45 mm, and a bandwidth of 2 GHz,
the rms noise level for a single baseline in a 5-min integration
is about 0.07/0.35 Jy. We require the calibrators to be strong
enough that the phase variations due to random noise fluctuations
in a given calibration scan are smaller than those produced by the
atmosphere between scans. For typical weather at Mauna Kea, over a
15 minute interval the atmospheric phase variations are 6 to 30
degrees for 10~m to 100~m baselines at 0.85 mm, and a factor of 2
higher at 0.45 mm \citep{mas94}. Such phase variations are
comparable with those seen in our observations of Uranus/Neptune
at 0.85 mm. For the SMA, the (antenna-based) rms phase noise is
28/X degrees \citep{bri89}, where X is the signal-to-noise ratio
for a single baseline. To achieve an rms phase precision of
$\sim$6 degrees over a 5-min calibrator scan, we require X to be
5, corresponding to a flux density of 0.35/1.8 Jy at 0.85/0.45 mm.
All four of the above-mentioned sources with point-like components
therefore qualify as calibrators at 0.85 mm, provided that short
baselines sensitive to the halo component are excluded and enough
baselines remain to derive (antenna-based) gain solutions.
Observations in superior weather conditions requiring higher phase
precision will naturally require stronger calibrators and/or
longer integration times.

For the above-mentioned sources to be suitable also as calibrators
at 0.45 mm, we require them to have a frequency spectrum
\emph{S}$_\nu$ $\propto$ $\nu$$^{>2.5}$. If the 0.85 mm emission
is produced entirely (or predominately) by dust, then we require a
dust opacity index $\beta$ $\gtrsim$ 0.5; such a dust opacity
index is commonly found in star-forming regions
\citep[e.g.,][]{mol00}. We therefore anticipate that
G15.04$-$0.68, IRAS~18089$-$1732, and IRAS 19217+1651, whose
cm-wave continuum emission is relatively weak ($\lesssim$30 mJy)
and hence their 0.85 mm emission dominated by dust, are suitable
as calibrators also at 0.45 mm. This, however, may not be the case
for G45.07+0.13. As shown in Figure \ref{g45-sed}, the flux
densities of G45.07+0.13 at 2$-$1.3 cm (15$-$22 GHz) measured with
the VLA range from 0.6 to 0.9 Jy, comparable to the measured flux
density of the core component at 0.85 mm. In this case, a large
fraction of the core emission may be produced by free-free
emission.

\subsection{Dust Distribution of UC \htwo Regions and HMPOs}

Three of the sources in Table~2 have been observed in the
continuum at 0.85~mm with the JCMT \citep{san94,jen95,wal03}.
Although the primary beam of the JCMT is a factor of 2.5 smaller
than that of the SMA, nevertheless the single-dish flux density is
a factor of 1$-$3 larger than that inferred for our core and halo
components combined. Because our measurements do not provide
stringent constraints on the radial intensity profile of the halo
component (modelled here as a symmetric Gaussian), we may have
underestimated its total flux density. For example, if the halo
component has a power-law radial density and/or temperature
profile instead of a Gaussian profile, its visibility amplitude
will decrease more rapidly with increasing {\it uv} distance and
hence its total flux density higher. On the other hand, there may
be a distinct, very extended ($\gtrsim$10$\arcsec$) component that
our interferometric observations do not detect. The halo
components that we observe have sizes ranging from 0.02 to 0.2 pc,
which is comparable to or smaller than the molecular envelopes
traced in CS (\emph{J}=5$-$4) around young high-mass stars of
$\sim$0.3 pc \citep{shi03}. Thus, the halo component may trace the
inner regions of this dense molecular envelope.

The presence of a core component implies a compact region that is
denser and/or warmer than its immediate surroundings.
\citet{van00} also find compact ($<$2\arcsec) dust components in
several HMPOs and UC \htwo regions at 3~mm and 1.3~mm, and suggest
that these components may originate from circumstellar disks.
Circumstellar disks are commonly seen around low-mass protostars,
but have been inferred only around a few massive young stellar
objects. \citet{yor96} postulate that the photoevaporation of such
disks is required to inject material to sustain UC \htwo regions.
For a dust temperature of 50~K (and $\beta$ = 1), the inferred
masses (for an interstellar gas to dust ratio of $\sim$100) of the
compact components are 3$-$50 \emph{M}$_\sun$. By comparison,
\citet{she01} reported a (dust) disk around G192.16$-$3.82 with a
mass of 3 \emph{M}$_\sun$, while \citet{zha98} reported a (NH$_3$)
disk around IRAS 20126+4104 with a minimum mass of $\sim$5
\emph{M}$_\sun$. Of course, a part or all of the core emission
could originate instead from a centrally collapsed region that
surrounds the central star. For UC \htwo regions, the compact
component could in addition be produced by warm dust either
cospatial with or in a relatively thin layer around the ionized
gas, as seen in the mid-IR \citep{kra01,kra03}.

\subsection{Future Directions}
In our initial search for calibrators among eight young high-mass
stars, we identified four potential candidates. Although
promising, we still need to demonstrate that calibration off the
point-source component in these sources can be successful (this
test was not possible here given the limited number of baselines
that sampled the point-source component). If not, we plan to
investigate whether detailed maps of these sources can be used as
an input for gain calibration (analogous to the case in VLBI
experiments). Bright and compact \htwo regions whose free-free
emission is by itself sufficiently strong to serve as calibrators
at sub-mm wavelengths (e.g., G45.07+0.13) should also be explored.

In its sparsest configuration, the SMA can attain an angular
resolution of $\sim$0\farcs3 at 0.85 mm and $\sim$0\farcs15 at
0.45 mm. Even if sufficiently compact dust sources associated with
high-mass star-forming regions can be found, it is unlikely that
they will be bright enough to serve as useful calibrators. To
illustrate, an opaque dust source with a temperature of $\sim$200
K (the highest so far measured in high-mass star-forming regions;
\citet{ces94}) and a size of 0\farcs3/0\farcs15 at 0.85/0.45 mm
has a flux density of $\sim$1.8 Jy. This is at the lower limit
required for calibration at 0.45 mm, and for more reasonable dust
temperatures and opacities, may not exceed the required flux
density to allow calibration even at 0.85 mm. Thus, for
observations at the highest angular resolutions with the SMA where
no known calibrators close to the target source exist, we may have
to resort to 2-D mapping of relatively compact ($\lesssim$
1$\arcsec$) dust sources (if such sources exist). Finally, the SMA
was designed to observe in two different wavelength bands
simultaneously to allow observations of a calibrator at a longer
wavelength to be used to correct the gain of the telescope at a
shorter wavelength, but this capability remains to be implemented
and tested.

\acknowledgments We thank all SMA staff members for their diligent
work in completing the SMA. J.L. and Y.S. thank the National
Science Council of Taiwan for support of this work through grant
NSC 92-2112-M-001-032. H.B. acknowledges financial support by the
Emmy-Noether-Program of the Deutsche Forschungsgemeinschaft (DFG,
grant BE2578/1).

\clearpage

\begin{deluxetable}{lccrcccll}
\tabletypesize{\scriptsize} \tablecaption{Target Sources}

 \tablewidth{0pt}
 \tablehead{\colhead{} & \multicolumn{2}{c}{Position(2000)}  & \colhead{S$_{2cm}$} &
 \colhead{2cm Size} & \colhead{S$_{1.3mm}$} & \colhead{S$_{350\mu m}$} & \colhead{Gain} & \colhead{} \\
            \colhead{Source} &
            \colhead{$\alpha$}  &
            \colhead{$\delta$} &
            \colhead{(mJy)} &
            \colhead{Maj$\arcsec$ $\times$ Min$\arcsec$} &
            \colhead{(Jy/(11\arcsec)$^2$)} &
            \colhead{(Jy)} &
            \colhead{Calibrator} &
            \colhead{\sc Ref.}
} \startdata
 UC \htwo \sc Regions&            &                &           &                  &   &        &          &     \\
 G10.841$-$2.592 &  18 19 12.12   & $-$20 47 31.3  &   8.7~    &   $<$0.5         &   &   320  & NRAO~530  & 1,2 \\
 G15.04$-$0.68   &  18 20 24.82   & $-$16 11 35.0  & 158.7~    & 0.4 $\times$ 0.3 &   &        & NRAO~530  & 3   \\
 G45.07+0.13     &  19 13 22.08   &   +10 50 53.4  & 594.2~    & 0.7 $\times$ 0.5 &   &   180  & 1925+211 \& 2015+371 & 3,4 \\
 Onsala 1        &  20 10 09.08   &   +31 31 35.7  & 203.9~    & 0.7 $\times$ 0.6 &   &   110  & 2015+371 & 1,2 \\
 G80.865+0.420   &  20 36 52.19   &   +41 36 24.3  &  40.0~    & 0.8 $\times$ 0.6 &   &        & 2015+371 & 1   \\
 \\
 HMPOs           &                &                &           &                  &   &        &          &     \\
 IRAS18089$-$1732&  18 11 51.46   & $-$17 31 28.8  &  0.9$^a$  &                  &2.1&  220   & NRAO~530  &4,5,6    \\
 IRAS18264$-$1152&  18 29 14.42   & $-$11 50 23.5  & $<$ 1$^a$ &                  &1.8&        & NRAO~530  &5,6      \\
 IRAS19217+1651  &  19 23 58.79   & $+$16 57 41.5  &   32$^a$  &                  &0.6&        & 1925+211 \& 2015+371 &5,6      \\
\enddata
\tablenotetext{a}{S$_{3.6cm}$} \tablenotetext{~}{{\sc{Reference}}
(1) \citet{kur94}; (2) \citet{hun00}; (3) \citet{woo89}; (4)
\citet{mue02}; (5) \citet{beu02}; (6) \citet{sri02}}
\end{deluxetable}

\clearpage

\begin{deluxetable}{lcccc}
\tabletypesize{\scriptsize} \tablecaption{Results of Visibility
Modeling}

 \tablewidth{0pt}
 \tablehead{\colhead{} & \colhead{$\it{uv}$ range} & \multicolumn{2}{c}{Halo Component}  & \colhead{Point-like Source}
 \\        \colhead{Source}           &
           \colhead{(k$\lambda$)}     &
           \colhead{Size (\arcsec)}   &
           \colhead{Amp$^a$ (Jy)}     &
           \colhead{Amp$^a$ (Jy)}
} \startdata
 G15.04$-$0.68   &  13 $-$ 80    &  9.56$\pm$0.71 &  14.18$\pm$3.14~~& 0.53$\pm$0.05   \\
 G45.07+0.13     &  20 $-$ 80    &  3.82$\pm$0.53 &  2.71$\pm$0.45  & 0.90$\pm$0.12    \\
 G80.865+0.420   &  18 $-$ 77    &  2.98$\pm$0.04 &  0.90$\pm$0.01  &  ---$^b$         \\
 IRAS18089$-$1732&  13 $-$ 81    & 2.68$\pm$0.17  &  4.63$\pm$0.25  & 1.72$\pm$0.20     \\
 IRAS18264$-$1152&  ~~27 $-$ 140 & 1.13$\pm$0.09  &  1.31$\pm$0.10  & ---$^b$           \\
 IRAS19217+1651  &  18 $-$ 80    & 1.99$\pm$0.17  &  0.92$\pm$0.07  &  0.35$\pm$0.04    \\
 \enddata
 \tablenotetext{a}{Not including the 25\% flux scale uncertainty}
 \tablenotetext{b}{One-dimensional Gaussian model is sufficient to represent the observed visibilities}
 \end{deluxetable}

 \clearpage
\begin{figure}

 \epsscale{0.9}
 \plotone{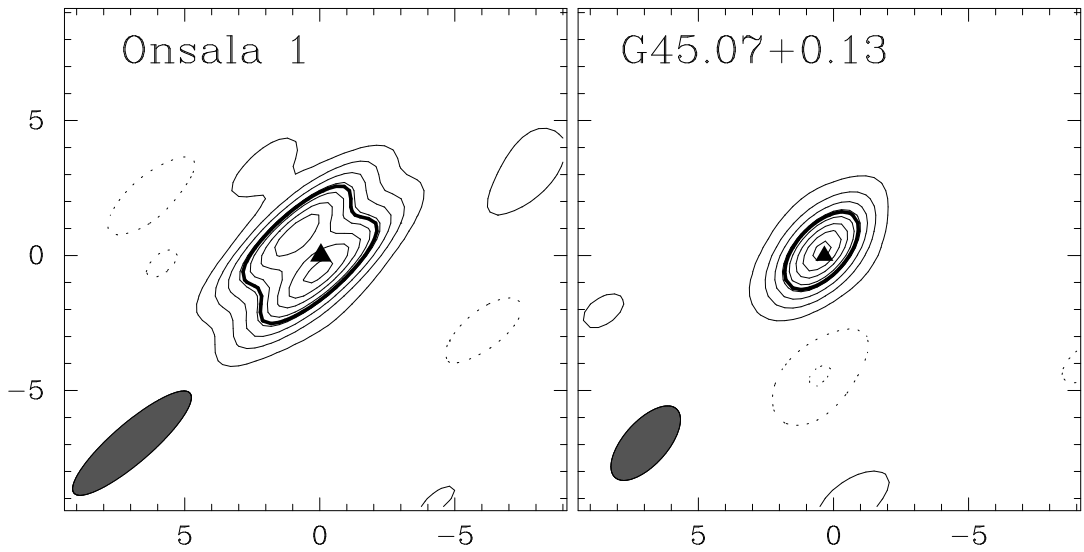}
 \vspace{-3cm}
 \caption{Contour plot of the 0.85 mm continuum emission imaged
towards Onsala~1 (left panel) and G45.07+0.13 (right panel). The
thick line is drawn at half the peak intensity. The cm-wave
continuum position is indicated by a triangle and the dark ellipse
denotes the synthesized beam. Contour levels are $-$0.09 (dotted),
0.09 to 0.93 in steps of 0.12 Jy beam$^{-1}$ in the left panel and
$-$0.28, $-$0.12 (dotted), 0.12 to 1.24 in steps of 0.16 Jy
beam$^{-1}$ in the right panel. \label{on1-map}}
\end{figure}

\clearpage

\begin{figure}
 \includegraphics[angle=270,scale=0.6]{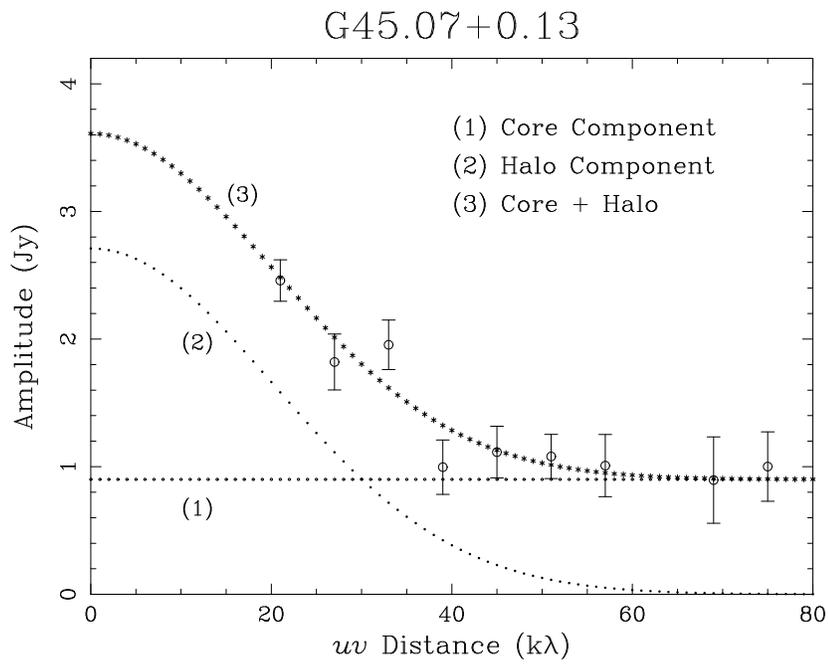}
 \caption{Plots (open circles) of measured visibility amplitudes versus {\it uv} distance averaged in circular annuli.
 Error bars indicate the $\pm$1$\sigma$ uncertainty in the visibility data for each bin. Model visibilities for a: (1) point source (core component); (2) one-dimensional Gaussian source (halo component); and (3) both sources together are shown by the dotted lines. \label{g45-uv}}
\end{figure}

\clearpage

\begin{figure}

 \includegraphics[angle=270,scale=0.6]{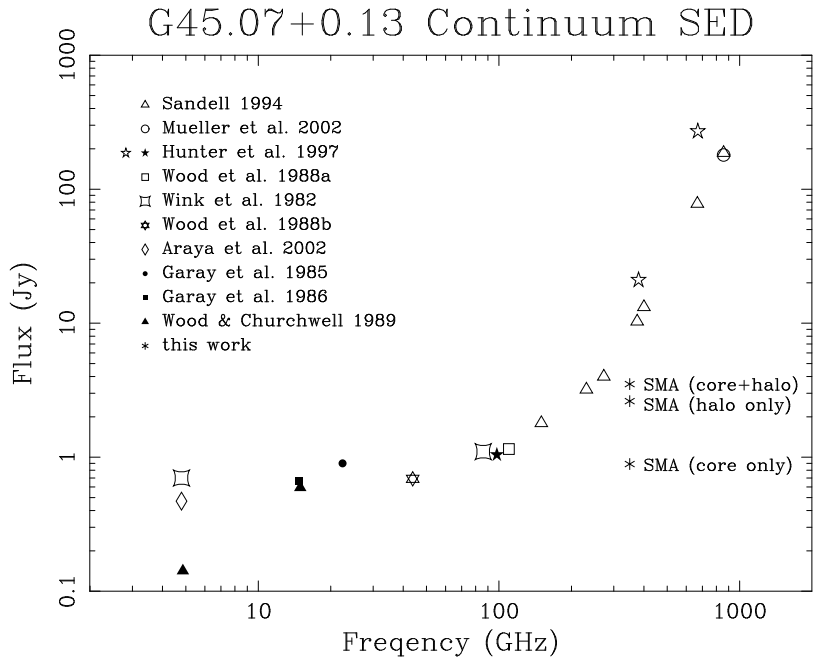}
 \caption{Spectrum of the continuum emission towards G45.07+0.13.
 The filled symbols represent measurements by interferometers such
 as the VLA, OVRO and SMA with typical angular resolutions of
 2\arcsec, while the open symbols represent measurements by
 single-dish telescopes with typical resolutions of 20\arcsec. \label{g45-sed}}
\end{figure}

\end{document}